# All passive photonic power divider with arbitrary split ratio

*Ke Xu*[\*†], *Lu Liu*[†], *Xiang Wen, Wenzhao Sun, Nan Zhang, Ningbo Yi, Shang Sun, Shumin Xiao*[\*], *and Qinghai Song*

Harbin Institute of Technology (Shenzhen), HIT Campus, The University Town of Shenzhen, Xili, Shenzhen, China, 518055.
[†]Equally Contribution.
[\*]Corresponding author: kxu@hitsz.edu.cn, shuminxiao@hitsz.edu.cn

**Integrated optical power splitter is one of the fundamental building blocks in photonic integrated circuits (PIC). Conventional multimode interferometer based power splitter is widely used as it has reasonable footprint and is easy to fabricate. However, it is challenging to realize arbitrary split ratio especially for multi-outputs. In this work, an ultra-compact power splitter with a QR code-like nanostructure is designed by a nonlinear fast search method (FSM). The highly functional structure is composed of a number of freely designed square pixels with the size of 120nm × 120nm which could be either dielectric or air. The lightwaves are scattered by a number of etched squares with optimized locations and the scattered waves superimpose at the outputs with the desired power ratio. We demonstrate 1×2 splitters with 1:1, 1:2, 1:3 split ratios and a 1×3 splitter with the ratio of 1:2:1. The footprint for all the devices is only 3.6µm×3.6 µm. Well-controlled split ratios are measured for all the cases. The measured transmission efficiencies of all the splitters are close to 80% over 30 nm wavelength range.**

## 1. INTRODUCTION

Photonic integrated circuits have become a new paradigm for on-chip subsystem with a plethora of applications like optical computing [1, 2], interconnection [3-6], sensing [7, 8] and so on. Similar to the electronic integrated circuits, the success of PIC origins from the large scale integration of various devices and functionalities on a single chip. As one of the most fundamental building blocks in PIC, the power splitter/combiner is widely used in many optical circuits. For the past years, the multimode interferometer (MMI) is a preferred candidate for this function due to its compactness, large optical bandwidth and excellent fabrication tolerance. Normally, a passive MMI divides the power equally to each output due to the self-imaging. However, arbitrary split ratio attracts more interests for many purposes such as signal monitoring, feedback circuits, power equalization and so on so forth. To realize this function based on MMI, the device structure either needs sophisticated modification to break the symmetry or requires the tuning scheme. Specially designed asymmetric MMI has been demonstrated with split ratio other than unity for 1×2 splitter [9, 10]. But it is difficult to design a multi-output splitter due to the increased design complexity. Adding electro-optic tuning element is also able to realize arbitrary split ratio and could be applied to multi-output device [11]. The cost will be increased under this case as the electro-optic tuning elements require more fabrication steps and the power consumption will be significantly larger. Directional coupler is an alternative while cascaded multi-stage coupler is needed for multi-output and the operation bandwidth is quite limited by the phase matching condition. Though much effort has been made, the existing design philosophy relies too much on the physical laws and design experience. Actually the design space is hardly fully explored under this case.

There are quite a few optimization methods developed to search the parameter space rather than sweeping and fine tuning a very limited number of parameters. Objective-first topology optimization [12] has been recently applied to design the wavelength multiplexer [13] and mode multiplexers [14]. Genetic algorithm was used to design a resonator [15] and grating coupler [16] as well. A similar algorithm relied on discretized pixel direct searching has been reported to design the polarization beam splitter [17], bending [18] and coupler [19] with very compact sizes. Recently, we also demonstrated a barcode-like waveguide-fiber coupler using a fast search method (FSM) which is capable of dealing with multi-objectives [20]. Thus far, all the above design strategies are based on dielectric permittivity engineering in a deep subwavelength scale and hence generate an optimum structure according to the specification. Since the automated electronic integrated circuits design has achieved a huge success, it is believed that such design automation would be promising and revolutionary for advanced photonic circuits.

In this paper, we have demonstrated a compact power divider/combiner using a QR code-like structure which could achieve arbitrary power split ratio for the first time. According to our best knowledge, this is the first all passive multi-output power splitter which could potentially achieve arbitrary split ratio. The device structure is optimized by FSM using a normal desktop. Compared with the topology optimization method, the advantage of this method is that the minimum device feature (pixel) size could be defined according to the nano-patterning resolution.

## 2. DEVICE DESIGN

Silicon photonics is one of the most promising platform for PIC due to a lot of remarkable advantages. Thus we design the device on the silicon on insulator (SOI) substrate with 220nm top silicon and 3μm buried oxide (BOX). The splitter is divided into 30×30 pixels and each of them is a square of 120×120 nm² which could be easily fabricated using the state of art lithography and dry etching technique. To this end, all the devices we are discussing below have the same size of 3.6×3.6μm². The pixels have a binary state of the dielectric property: silicon or air. We start from an all silicon structure which is a normal MMI to initialize the optimization process depicted as Fig. 1 (a). We switch one pixel state in each iteration to determine whether the pixel state should be silicon or air. Hence the refractive index could be engineered at every location of the device in a deep subwavelength scale. Then a few figure of merits (FOM) are defined and numerically calculated by 3D finite difference time domain (FDTD) simulation. They are the transmission efficiency at each output and the power ratios between different outputs. Specifically, the TE mode light is launched into the input single mode waveguide and the monitors are used to measure the power at the output waveguides. The algorithm keeps searching the structure that the FOMs are above a certain value. The values of FOMs are our objectives which have been defined prior to the simulation. The simulation iterates until the convergence to the objectives. Rather than simply increasing the power at all the output ports, the power ratios should be strictly controlled by different convergence conditions. Thus it becomes a multi-objective problem. If both the two FOMs are improved, the state of the pixel is saved. Otherwise, the state is reversed and the program proceeds to the next pixel. The convergence conditions in our algorithm are described by:

$$\left(\frac{\sum_i E_{i,j+1}}{\sum_i E_{i,j}} > 1\right) \cap \left[\left(\frac{E_{i,j+1}}{E_{i-1,j+1}} - \alpha\right)^2 < \rho_{j+1}^2\right]$$

where $E_{i,j}$ stands for the transmission efficiency of the i-th output waveguide in the j-th iteration. We define the power split ratio as α. Here we apply an FSM to speed up the parameter searching which could be referred to Ref. [20]. Since fast rate of convergence usually requires a strong convexity of the objective [21], an elliptical equation (a function of $E$ and α) is used to approximate the radius of convergence $\rho_j$ when the transmission efficiency is below a certain threshold. Afterwards, a linear function is used to approximate the $\rho_j$ when the efficiency is higher. Finally, the device is optimized for a wide range of wavelengths to ensure a broad band operation. The optimized structure has a number of randomly located square holes and the schematic picture of a 1×3 splitter is shown by Fig. 1 (b).

We use a normal 4-core desktop to design the device by automatically running the program. The computation cost is ~120 hours in average to get the optimized results. The simulated electric fields at a wavelength of 1550nm are plotted in Fig.(2). The power dividing mechanism for this type structure is different from the conventional MMI splitter. The launched electromagnetic waves reaches the randomly located scattering center and the scattered waves superimpose at the output waveguides. The desired power split ratio could be achieved by a proper structure. We first design three 1×2 power splitters with three different ratios of 1:1 (device 1), 1:1.5 (device 2) and 1:2 (device 3) to show the possibility of precise split ratio control. The simulated electric fields for the optimized structures are shown in Fig.2. (a)-(c) respectively. Compared with the 1×2 splitter, the realization of a multi-output power splitter with arbitrary power ratio would be more attractive. The design method in this work could be applied to a more general case with arbitrary split ratio and arbitrary number of output ports in principle. Here we show a proof of concept demonstration of a 1×3 splitter with splitting ratio of 1:2:1 (device 4). The electric field distribution for the optimum structure is simulated and shown in Fig. 2. (d). A higher intensity could be observed at the central output waveguide than the other two ports. The simulated power split ratios over a wide wavelength range are shown in Fig.3 (a), (b) for the 1×2 and 1×3 splitters respectively. A negligible error of the split ratio over the wavelength range from 1530nm to 1560nm is observed. The algorithm we developed has a good control on the split ratios with acceptable errors for the 1×2 splitters over 30nm wavelength range. Here we have not considered the operation bandwidth as an objective. It could be improved further for wide band operation while the computation cost will be higher at the same time by setting another objective. The splitting ratio of the 1×3 splitter is plotted in Fig. 3 (b) where the red line indicates the ratio between the central output port and the lower output port. The blue line describes the ratio between the central port and the upper output port. Both the two ratios equal to 2:1 as we expect. But there is a small discrepancy at longer wavelength. This could be improved if we increase the optimization time. It should be noted that the simulated transmission efficiencies of all the devices are above 80% within the whole spectrum range from 1530nm to 1560nm. The excess loss is expected to mainly come from the scattering of each etched holes.

### 3. EXPERIMENTAL RESULTS

The above devices are fabricated on a commercial SOI wafer (SOITEC Inc.) with 220nm top silicon and 3μm BOX. The device layout is patterned by electron beam lithography (Raith eLINE) which operates at 30kV. The positive photoresist ZEP 520A is used as the mask. The pattern is then transferred in device layer using one step inductively coupled plasma (ICP) etch process. The power splitter has a small footprint of only 3.6×3.6μm². The schematics of each optimized structure are shown in Fig. 4. (a)-(d) which are corresponding to device 1-4. The white square indicates the etched air holes and they are randomly located within the device area. The top-view scanning electron microscope (SEM) images of the fabricated devices are shown in Fig. 4. (e)-(h). Though some of the squares have round corners due to the imperfection of the fabrication process, the following experimental results actually indicate a good robustness of the designs.

The experiment is carried out to evaluate the device functionality and performance. Continuous wave (CW) light from a tunable laser is coupled into the submicron waveguide through a subwavelength focusing grating coupler with 10 degrees tilt. The coupling loss is measured to be ~ 7dB per grating at the peak wavelength. It is a bit higher than the reported loss and this is due to a single deep etch process is used in our fabrication. It could be improved further by non-uniform period or shallow etched structure. The output power at each port is then measured by coupling out the light from the chip with another grating coupler and received by a photodetector. The top-view SEM of a whole 1×3 splitter is shown in Fig. 5. (a). The 1×2 power splitter is characterized by scanning the wavelength and recording the output power ratio between each port. The measured results of device1-3 are plotted in Fig. 5. (b). Flat responses for the three devices are obtained over 30 nm spectrum range. The measured results are well consistent with the objectives we set. The device1 and 3 have a larger error at short wavelength near 1530nm but this agree with the predictions from simulations. The power split ratio of the 1×3 splitter (device4) is measured and is shown in Fig. 5. (c). It depicts the power ratio between the central output and the upper/lower output which should be both 2:1. The measured transmission efficiencies of device1-4 are quite close to 80%. By optimizing the etching recipe, we believe these values could reach to the theoretical values. Thus far, we have successfully demonstrated a series of power splitters with the target split ratios.

On the other hand, the optimum transmission efficiency of a conventional MMI splitter is highly dependent on the splitter geometry especially the length of the multimode region. However, our method has realized both 1×2 and 1×3 splitters with different split ratios and the same size which are hardly possible for normal MMI. This is a significant promise to broaden the current device library developed by the formal design method and to establish a new device design fashion. To validate the contributions from the randomly located squares, we have done a control experiment by fabricating a conventional all-silicon MMI. The all-silicon MMI has the same geometry as device4 ($3.6 \times 3.6 \mu m^2$) and has three output ports. The top-view SEM image of the all silicon normal MMI is shown in the inset of Fig. 6. We also measured the transmission efficiencies of each port and the split ratio. The measured transmission efficiencies of output port 1-3 are plotted in Fig. 6. It turns out that the efficiencies are quite low without the pixels that are etched. The central output port has nearly 20% efficiency while the other two ports have less than 5% efficiency. And the split ratio is far from what we designed. In this sense, our optimization algorithm works well and the QR code like nanostructure is found to be highly efficient and functional.

## 4. CONCLUSION

We have experimentally demonstrated a potential technique to achieve integrated optical power splitter with arbitrary ratio. The device is optimized by a fast searching method with reasonable computing cost. According to our best knowledge, it is the first demonstration of a passive silicon power splitter with multi-output ports which are optimized by algorithm. We have shown three 1×2 splitters with ratio of 1:1, 1:2 and 1:3. To demonstrate the scalability for multi-output operation, we also report a 1×3 splitter with ratio of 1:2:1. The experimental results are consistent with the simulation results and have negligible errors from the design targets. This work is a nice example to demonstrate the possibility of the design automation for realization of highly functional device that is hardly possible for conventional first-principle design philosophy.

**Funding**. National Natural Science Foundation of China (NSFC) (61505039); Shenzhen Municipal Science and Technology Plan Project (JCYJ20150403161923530).

## Figures

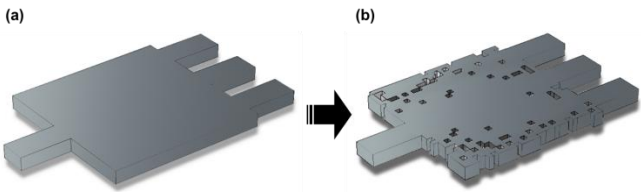

Fig. 1. (a) The schematic picture of the normal all silicon MMI structure for initialization. (b) The schematic picture of the optimized structure.

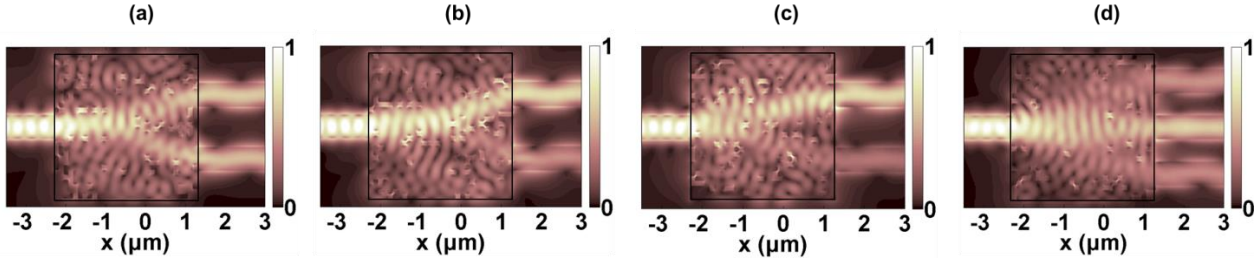

Fig. 2. (a)-(d) The simulated electric field for the optimized device1-4 respectively.

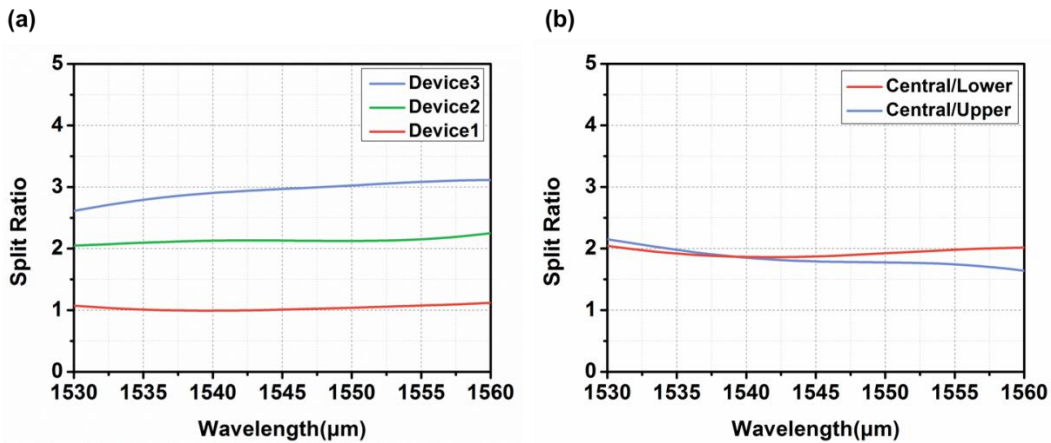

Fig. 3. (a) The simulated split ratios for the 1×2 splitters (devices 1-3). (b) The simulated power split ratio for the 1×3 splitter (device4) where the blue line indicates the simulated power ratio between the central port and the upper port. The red line indicates the ratio between the central port and the lower port.

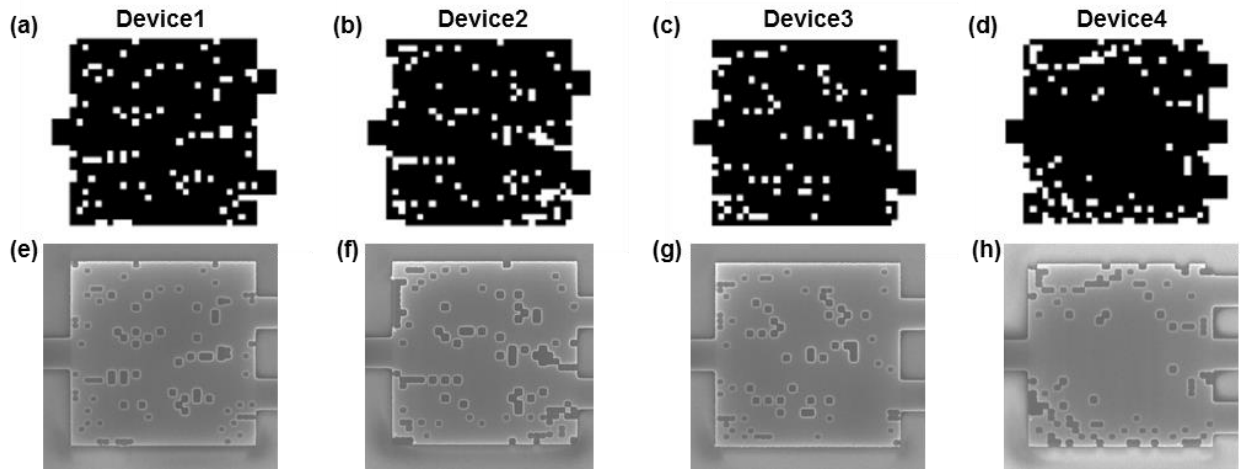

Fig. 4. (a)-(d): The QR code like device geometry for device1-4 with the power split ratio of 1:1, 1:2, 1:3 and 1:2:1, respectively. (e)-(h): The SEM images of the device1-4 correspondingly.

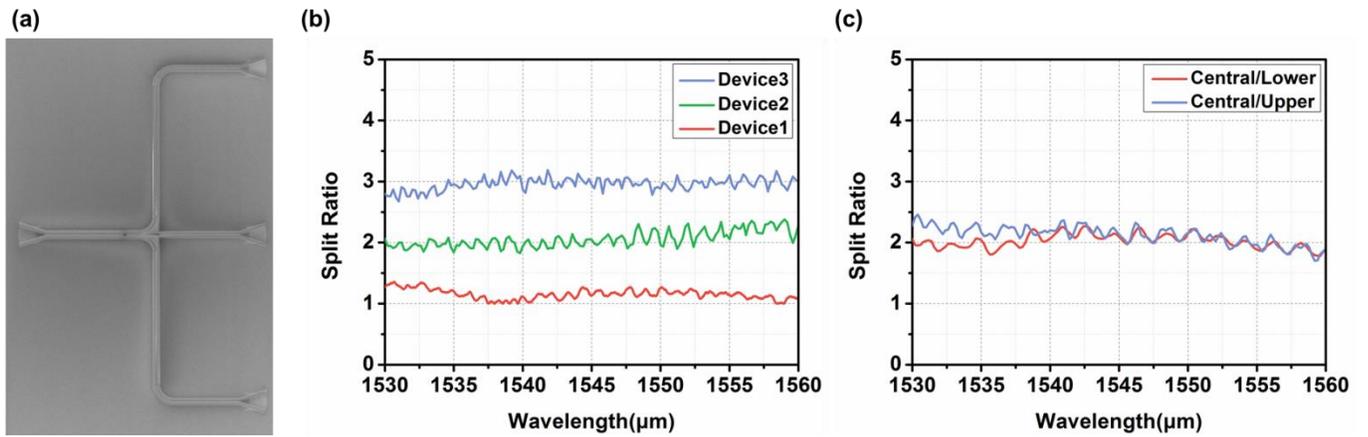

Fig. 5. (a) The top-view SEM of the whole device. (b) The measured split ratio for the 1×2 splitter (device 1-3). (c) The measured power split ratio for the 1×3 splitter (device4). The blue line indicates the measured power ratio between the central port and the upper port. The red line indicates the ratio between the central port and the lower port.

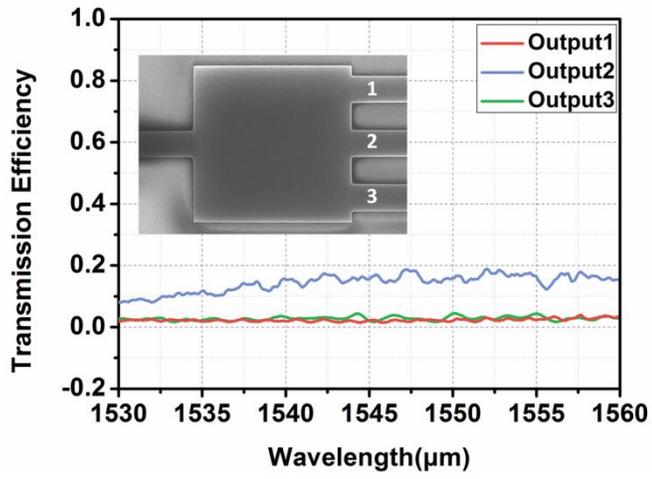

Fig. 6. The measured transmission efficiencies of output port 1-3 of the all silicon conventional MMI. Inset: the top-view SEM image of the device.